\documentclass[a4paper]{jpconf}
\usepackage{graphicx}
\usepackage[sort&compress,numbers]{natbib}
\begin{document}
\title{From dripline to dripline: \\
Nuclear astrophysics in the laboratory}

\author{Zach Meisel}

\address{Department of Physics and Joint Institute for Nuclear
Astrophysics, \\ University of Notre Dame, Notre Dame, 46556 Indiana,
USA}

\ead{zmeisel@nd.edu}

\begin{abstract}
For the better part of a century the field of nuclear astrophysics
has aimed to answer fundamental questions about nature, such as the
origin of the elements and the behavior of high-density,
low-temperature matter. Sustained and concerted efforts in nuclear
experiment have been key to achieving progress in these areas and
will continue to be so. Here I will briefly review recent
accomplishments and open questions in experimental nuclear
astrophysics.
\end{abstract}

\section{Nearly 100 years of nuclear astrophysics}
\label{intro}
\begin{figure}[ht]
\begin{center}
\includegraphics[width=0.8\columnwidth,angle=0]{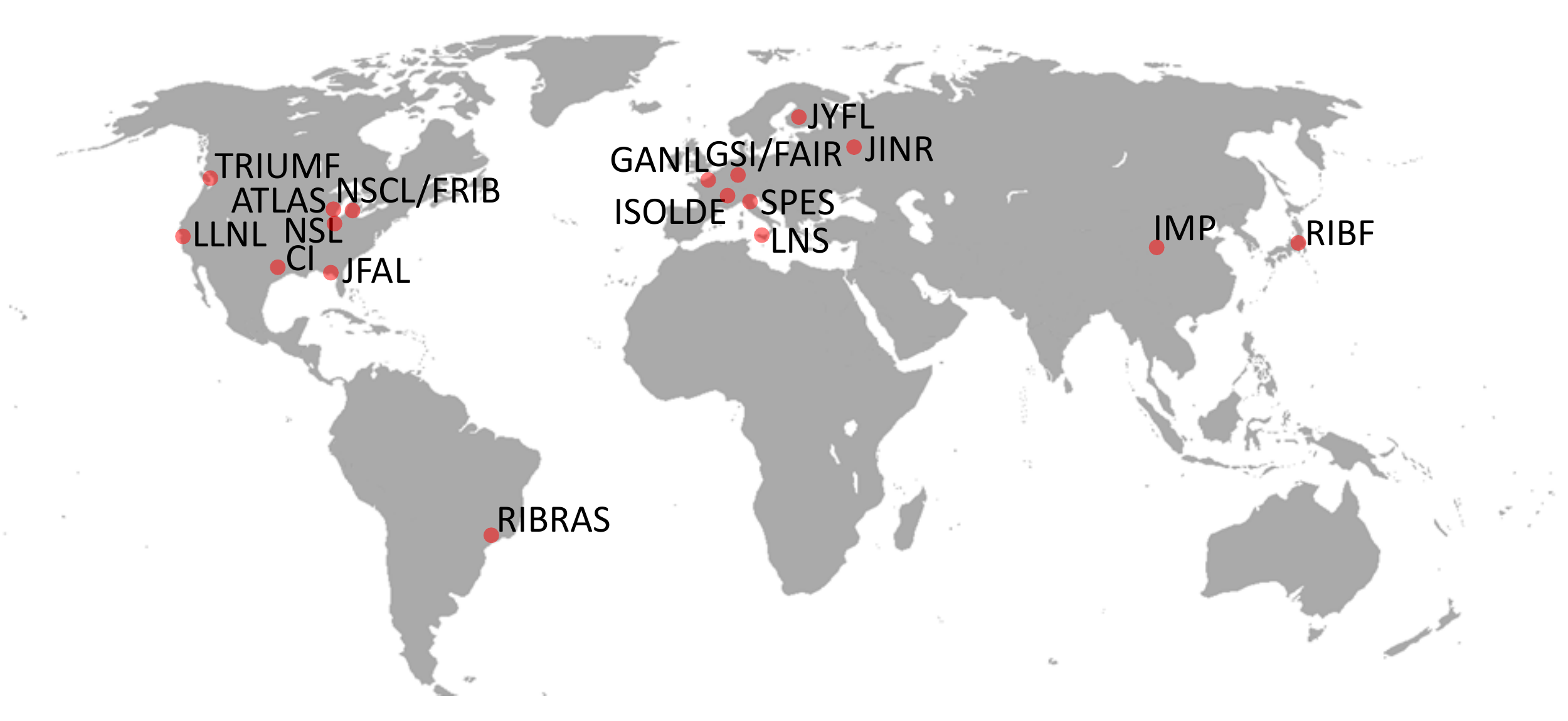}
\caption{
 Global census of radioactive ion beam facilities.
\label{GlobalRIB}}
\end{center}
\end{figure}

With Eddington's brilliant conjecture in 1920 that the Sun's energy
reservoir ``can scarcely be other than sub-atomic energy", based on
ground-breaking experimental work by Aston and Rutherford, the field
of nuclear astrophysics was born~\cite{Eddi20}.
Work accomplished in the following decades demonstrated that nuclear
physics was not only key to describing how stars lived and died, but
also how they made the elements we see (and are made of)
today~\cite{Burb57,Came57}.
To date, great strides have been made in a worldwide effort (See
Fig.~\ref{GlobalRIB}.) toward answering fundamental questions about
our universe, chief among them: \emph{Where were the elements
made?}, \emph{How does nuclear energy generation impact stars and
stellar explosions?}, and \emph{How does matter behave at high
density and low temperature?}. This progress has relied on the study
of atomic nuclei over nearly the entire nuclear landscape, the
region bounded by the so-called proton and neutron driplines, where
protons or neutrons `drip' out of the nucleus due to the extreme
mismatch in their respective numbers.

The dramatic enhancements of
experimental capabilities offered by next generation
radioactive ion beam facilities such as the Facility for Rare
Isotope Beams (FRIB)~\cite{FRIB} and the NuSTAR experiments at the Facility for Antiproton and
Ion Research (FAIR)~\cite{FAIR}, coupled with advances in
observational and computational capabilities
(e.g. Refs.~\cite{NUSTAR,UNEDF}), are certain to deepen our
understanding of nature and likely yield more than a few surprises.
The following sections will briefly touch on recent accomplishments in
experimental nuclear astrophysics and
outstanding questions across the nuclear landscape. Due to
space limitations, many exciting works and research topics have been
omitted. For more comprehensive reviews, see
Refs.~\cite{Jose11,Wies12,Scha16}.

\section{Progress on the proton-rich side}
\label{prich}
Various astrophysical environments feature nuclear reactions on the
neutron-deficient side of the valley of $\beta$-stability that are
responsible for nuclear energy generation and element formation.
These proton-rich conditions, caused by an injection of large amounts
of hydrogen from a stellar envelope or transmutation of neutrons
into protons via neutrino capture, enable sequences of proton-capture reactions
to proceed at high-temperature. These reactions drive the rapid proton-capture
($rp$)-process that largely powers type-I x-ray bursts~\cite{Pari13} and
classical novae~\cite{Bode12} and are a main player in the reaction
network leading to nucleosynthesis via the neutrino-p ($\nu
p$)-process in core-collapse supernovae~\cite{Froh06}.

Classical novae are thermonuclear explosions on the surfaces of white
dwarf stars that recur due to the reaccumulation of hydrogen fuel
from a binary companion star~\cite{Bode12}. These explosions, aside from
generating astronomical displays occasionally visible to the naked
eye, contribute to the creation of light elements in the universe.
Even though these objects have been a focus of intense study for
decades, including numerous observational, theoretical, and
experimental efforts, their exact contribution
to the cosmic abundances is still unknown. Major advances
have been made recently by using the power of pure, exotic
radioactive ion beams to investigate the origins of 
presolar dust grains that may have been produced by novae (e.g.
Ref.~\cite{Benn16}).
In fact, studies such as the aforementioned work make novae one of
the few astrophysical phenomena for which most nuclear
reaction rates are based on experimental data.

Type-I x-ray bursts are similar but much more powerful explosions
recurring on the surfaces of hydrogen and helium accreting neutron
stars, where the larger surface gravity of the neutron star relative
to the white dwarf ultimately leads to the enhanced energy
release~\cite{Pari13}. These observables are one of the main
tools used for understanding neutron stars, unique astronomical
laboratories that provide insight into the behavior of dense matter at
low temperature. Recent work has focused on understanding reactions
that trigger the $rp$-process (e.g. Ref.~\cite{Parp05}) and the
locations of the nuclear landscape where the $rp$-process reaction
sequence is significantly stalled, termed waiting-point nuclides (e.g.
Ref.~\cite{DelS14}). Though the strengths of all waiting-points are
soon to be well constrained, theoretical work has shown that a host
of other nuclear uncertainties remain that must ultimately be
removed or reduced by future experiments~\cite{Cybu16}.

\section{New and old directions for neutron-rich nuclides}
\label{nrich}
On the opposite side of the nuclear chart, an array of nuclear
reaction sequences operate in astronomical environments to various
ends, such as the forging of new elements and the alteration of
dense objects' thermal and compositional structures. The oldest and most well known
of these is the rapid neutron-capture ($r$)-process that is
responsible for creating roughly half of the elements heavier than
iron in as-yet undetermined astrophysical sites~\cite{Mump16}.
However, other reaction sequences such as the
$\alpha$-process~\cite{Arco11} and $i$-process~\cite{Jone16} have
recently joined the club of potential mechanisms for nucleosynthesis
on the neutron-rich side of stability. Equally exciting are the new
developments that have shown individual nuclear properties are
critical to understanding the various observables from neutron stars
that provide unique windows into the behavior of high-density,
low-temperature matter~\cite{Scha14}.

\begin{figure}[ht]
\begin{center}
\includegraphics[width=0.9\columnwidth,angle=0]{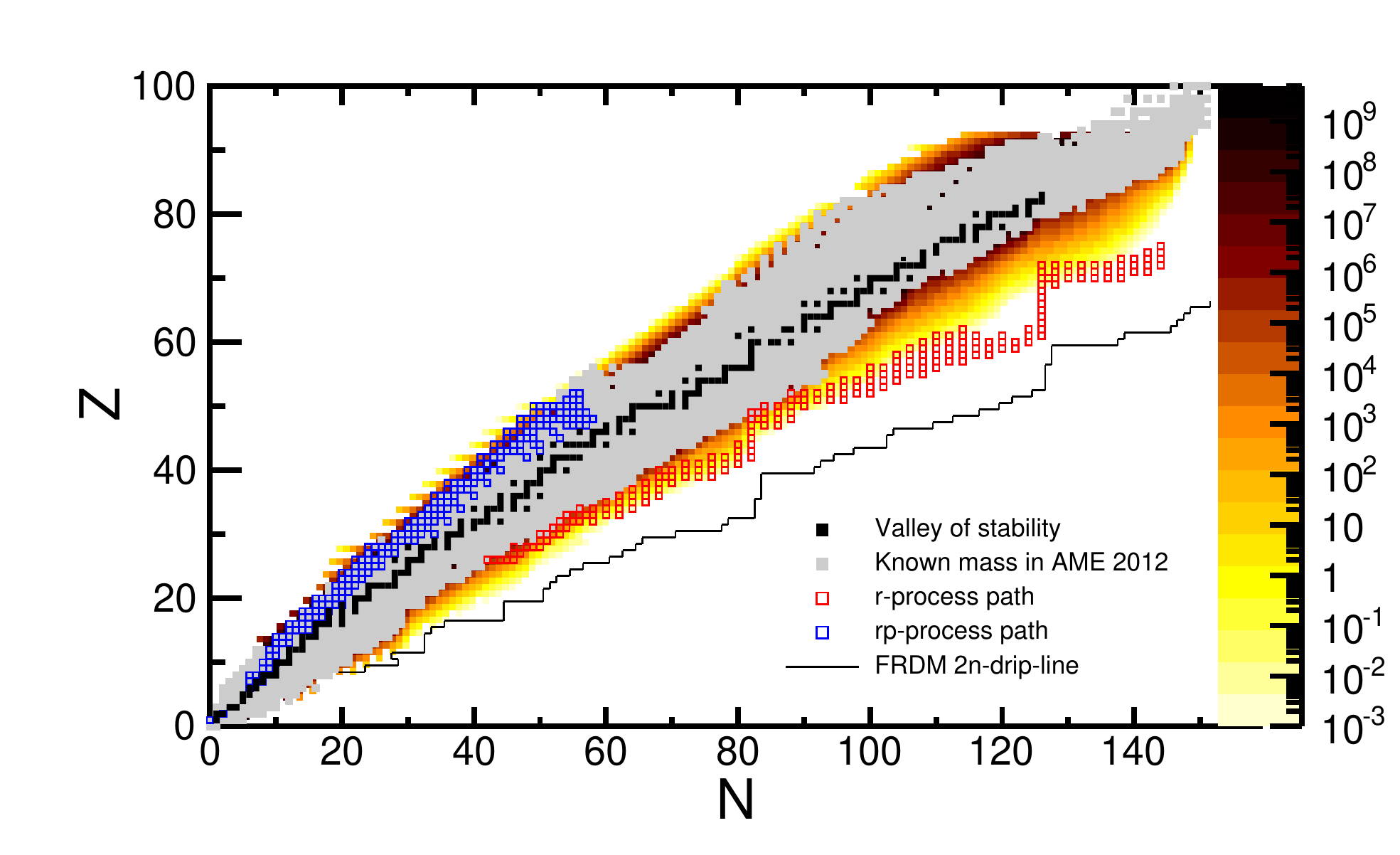}
\caption{
  Predicted FRIB production rates in particles per
  second~\cite{Boll11}. See Ref.~\cite{Fran08} for a similar
  prediction for FAIR.
\label{FRIBproduction}}
\end{center}
\end{figure}

The $r$-process, the rapid neutron-capture sequence likely operating
in core-collapse supernovae and/or neutron star mergers, has long
been out of reach of even the most advanced radioactive ion beam
facilities. To date, much of the relevant experimental work has focused on
constraining key nuclear quantities nearer to stability so that
these results can guide theoretical estimates for nuclides on the
$r$-process path~\cite{Mump16}. However, recent innovative
techniques have allowed some of the first experimental constraints
to be made for nuclear reactions on the path itself (e.g.
Ref.~\cite{Spyr14}).
This and other approaches will be particularly powerful when coupled
with the extended reach toward the neutron dripline anticipated for
FRIB (See Fig.~\ref{FRIBproduction}.) and FAIR and promise to dramatically advance our understanding
of this long-studied problem.

In the past decade other avenues for nucleosynthesis have been
identified that likely operate nearer to stability on the
neutron-rich side. The $\alpha$-process, which may operate via a
sequence of $\alpha$-capture, neutron-emission
reactions~\cite{Peri14}, and
$i$-process, a reaction sequence similar in spirit to the
$r$-process but at lower neutron densities, have only just begun to
be investigated. Numerous experimental studies are anticipated in the near future
that will drastically expand our knowledge of these processes and provide
critical tests of their viability as mechanisms of element
formation.

Lately it has been shown that neutron-rich nuclides are
just as important as their proton-rich cousins with regards to their
impact on our understanding of dense matter. When the proton-rich
ashes of the $rp$-process are compressed on the neutron star surface
by subsequent accretion from a binary companion, electrons are
forced into nuclei, converting protons into neutrons and
fundamentally altering the neutron star thermal and compositional
outer structure~\cite{Scha14}. It has been shown that it is
critical to determine the properties of individual neutron-rich
nuclides in order to accurately describe the accreted neutron star
ocean and crust~\cite{Estr11,Meis15,Meis16}. Efforts in the near
future are planned to focus on nuclides which have been identified
to have the greatest potential impact on astronomical observables~\cite{Deib16}.

\section{Nuclear astrophysics near stability}
\label{stable}
Although most nuclides in and near the valley of $\beta$-stability
have been accessible in the laboratory for some time, several
outstanding questions in nuclear astrophysics require their further
study. These involve quiescent and explosive stellar environments and require
experimental approaches using both indirect and direct techniques to
study the most important reactions~\cite{Brun15}. Open questions
include the extent to which photodisintegration reactions impact the
cosmic elemental abundances~\cite{Raus13}, the exact abundance yield
from slow neutron capture in stellar envelopes~\cite{Reif14}, and
the role of electron captures in high-density astrophysical
environments~\cite{Sull16}. Progress in these areas has been driven
by several nuclear physics labs around the world, especially the
many stable-ion beam facilities which are far too numerous to
include in Fig.~\ref{GlobalRIB}.

Precisely describing the nuclear reaction sequences of stars has
remained a challenge since Eddington first approached the
subject~\cite{Eddi20}.
Partially due to triumphs of astrophysical modeling and observations, such as
asteroseismology and measurements of neutrinos from our sun, high
precision experimental studies are needed to advance our
understanding of quiescent nuclear burning in stars~\cite{Brun15}.
Specialized equipment such as recoil mass separators and underground laboratories (e.g.
Refs.\cite{Coud08} and \cite{Robe16}, respectively) have
played and will continue to play a major role in this effort. When
direct measurements via these and other methods are not possible, as is the
case for some branch-point nuclides in the slow neutron-capture
($s$)-process reaction network, indirect techniques will be required
to experimentally constrain important reactions~\cite{Reif14}.

In spite of their association with the most exotic nuclides, models
of stellar explosions require a thorough understanding of nearly the
entire nuclear landscape, including nuclides along and near
stability. A case-in-point is the photodisintegration-driven
$p$-process operating in supernovae, which is currently the favored creation
mechanism of the so-called $p$-nuclides whose
origins cannot
be explained by the $s$ and $r$ processes~\cite{Raus13}. Sustained
efforts have reduced the nuclear physics
uncertainties of this process, where the focus has generally been on
constraining the Wolfenstein-Hauser-Feshbach reaction theory that
provides essential input to astrophysics models in the absence of
experimental data (e.g. Refs.~\cite{Quin15,Yalc15}). Additional
measurements on and near stability have focused on reducing the
uncertainties in nuclear weak rates that limit the ability to
describe the mechanisms through which supernovae operate (e.g.
Ref.~\cite{Noji14}). Here theory calculations have provided
important guidance, identifying the most essential nuclear data and
filling in the large gaps left by insufficient experimental
information~\cite{Sull16}.

%\begin{figure}[ht]
%\begin{center}
%\includegraphics[width=0.75\columnwidth,angle=0]{ExoticityAstroProcesses.pdf}
%\caption{
% Census of known and unknown masses by exoticity as of
% 2012~\cite{Meis13}.
%\label{Exoticity}}
%\end{center}
%\end{figure}

\section{FRIB, FAIR, and the future}
\label{outlook}
Roughly 100 years after its inception, nuclear astrophysics
research continues to enhance our understanding of nature. At
present the field is poised to build upon our current body of
knowledge by leaps and bounds, in no small part due to upcoming developments such
as new recoil separators~\cite{Coud08,Meis16b}, underground
laboratories~\cite{Robe16}, and storage rings dedicated to nuclear
physics studies~\cite{Fran08}. Frontier nuclear physics facilities
such as FRIB and the NuSTAR experiments at FAIR will play a central role in this advancement
by providing unprecedented access to ever more exotic nuclides (See
Fig.~\ref{FRIBproduction}.). Meanwhile, stable beam facilities will
continue to play a complementary role in answering astrophysical
questions both new and old. In the near future, together with advances in observation
and theory, experimental nuclear astrophysics studies from dripline to dripline
promise to offer profound insight into how our universe operates.

\section*{Acknowledgements}
%I thank my collaborators at the Nuclear Science Laboratory at Notre Dame.
This work was supported by the National Science Foundation Grants No. 1419765 and 1430152.

%\clearpage

\bibliographystyle{iopart-num}
\bibliography{FairnessReferences}

\end{document}